\documentclass[aps,pra,twocolumn]{revtex4}

\usepackage{amsmath,amssymb}
\usepackage{times}
\usepackage{acronym,array,bm,color,dcolumn}
\usepackage{epsfig,graphicx,nomencl,upgreek,url}
\usepackage{multirow}
\usepackage[colorlinks={true},pdftex]{hyperref}  
\hypersetup{
  colorlinks,
  citecolor=green,
  linkcolor=blue,
  urlcolor=blue}

\makeatletter

\newcommand{\ket}[1]{\ensuremath{\left\vert{#1}\right\rangle}}

\newcommand{\expect}[1]{\ensuremath{\left\langle{#1}\right\rangle}}

\newcommand{\trans}[1]{#1^\mathsf{T}}

\renewcommand{\Im}{\operatorname{Im}}

\newcommand{\erf}[1]{Eq.~(\ref{#1})}

\newcommand{\beq}{\begin{equation}}
\newcommand{\eeq}{\end{equation}}
\newcommand{\bqa}{\begin{eqnarray}}
\newcommand{\eqa}{\end{eqnarray}}
\newcommand{\nn}{\nonumber}
\newcommand{\eg}{\emph{e.g.},~}
\newcommand{\ie}{\emph{i.e.},~}

\usepackage{theorem}
\theorembodyfont{\rmfamily}

\def\R{\mathbb{R}} 
\def\C{\mathbb{C}}

\def\CalL{\mathcal{L}}

\def\su{\mathfrak{su}}

\def\la{\langle}
\def\ra{\rangle}
\def\Eq{Eq.~\eqref}
\def\Eqs{Eqs.~\eqref}

\DeclareMathOperator{\tr}{tr}

\newcommand{\ma}[1]{\left[\begin{matrix} #1 \end{matrix}\right]}
\newcommand{\mb}[1]{\begin{matrix} #1 \end{matrix}}

\newcommand{\dg}{^{\dagger}}
\newcommand{\trp}{^{\sf T}}

\def\mA{\mathbf{A}}
\def\mG{\mathbf{G}}
\def\mb{\mathbf{b}}
\def\mx{\mathbf{x}}
\def\mC{\mathbf{c}}
\def\my{\mathbf{y}}

\def\mI{\mathbf{I}}
\def\zero{\mathbf{0}}
\def\bnu{\boldsymbol{\nu}}
\def\mH{\mathbf{H}_{rs}}

\newcommand{\SJTU}{Joint Institute of UMich-SJTU and Key Laboratory of
  System Control and Information Processing (MOE),
  Shanghai Jiao Tong University, Shanghai, 200240, China}
\newcommand{\SNL}{Department of Scalable \& Secure Systems Research (08961),
Sandia National Laboratories, Livermore, CA 94550, USA}

\begin{document}

\title{Identification of open quantum systems from observable time traces}

\author{Jun Zhang}
\affiliation{\SJTU}
\author{Mohan Sarovar}
\email[Electronic address: ]{mnsarov@sandia.gov}
\affiliation{\SNL}

\date{\today}

\begin{abstract}
  Estimating the parameters that dictate the dynamics of a quantum
  system is an important task for quantum information processing and
  quantum metrology, as well as fundamental physics. In this paper we
  develop a method for parameter estimation for Markovian open quantum
  systems using a temporal record of measurements on the system.
  The method is based on system realization theory and is a generalization of our previous work on identification of Hamiltonian parameters [Phys.
  Rev. Lett. 113, 080401 (2014)].
\end{abstract}

\pacs{03.65.Wj, 02.30.Yy, 03.67.-a}

\maketitle

\section{Introduction}
Recent years have witnessed the rapid progress of quantum information
processing technologies, including convincing demonstrations of a
\emph{quantum advantage} in several applications including
communication and sensing. Such technologies require the precise
fabrication and manipulation of quantum degrees of freedom, and as a
result, much effort is invested into understanding and precisely
identifying the quantum dynamics.

In Ref. \cite{JunMohan:14} we developed a technique for identifying a
parameterized Hamiltonian from time traces of expectation values of a
small set of observables. This technique was recently experimentally
demonstrated and validated in Ref. \cite{Hou:2014wg}. In the current
paper we generalize this work to enable identification of
parameterized open system evolution that can be described by a Lindblad
master equation. This expands the applicability of this type
of system identification approach that utilizes time traces of
observables. As in Ref. \cite{JunMohan:14} we consider only finite dimensional systems, and assume that the
system can be reliably prepared in a small number of initial states,
and possesses observables whose expectation value can be sampled over a
period of time. Nuclear magnetic resonance
\cite{Sli-1996, Hou:2014wg} and ensembles of neutral atoms \cite{Smith:2013gw}
are two typical examples of physical systems for which these
assumptions are valid.

Our approach can be considered a generalization of traditional
spectroscopic methods such as Ramsey interferometry in which
spectral features of time-dependent data are used to infer values of
underlying system parameters \cite{Sli-1996, Ger.Kni-2005}. This inference is
simple in the case of Ramsey or Rabi measurements where the relation
between spectral features and the parameters is straightforward. In
more complex situations, the relationship 
can be too complex to know \textit{a priori}.  Moreover,
it may not even be known whether the measurements performed \emph{can}
identify a parameter of interest.  Such complex situations beyond
conventional spectroscopy can occur even in small systems such as a
few atoms or spins. We develop a systematic way to perform parameter
estimation in such complex situations, providing a criterion of
whether the unknown parameters are estimable given the set of
measurements available, and if so, prescribing a data-driven algorithm
to identify them. 

In addition to Ref. \cite{JunMohan:14}, previous studies that have examined this kind of \emph{generalized spectroscopy} to estimate Hamiltonian or open system parameters from time traces include Refs. \cite{Boulant:2003ih, Col.Sch.etal-2005, Devitt:06, Burgarth:2009ee, Burgarth:2009dl, DiFranco:2009ju, Bellomo:2010bg, Bellomo:2010ef, Granade:2012kj, Schirmer:1452667, Kato:2013ub, Dominy:2013tk, Jagadish:2014ua}. We draw particular attention to Ref. \cite{Schirmer:1452667}, which studied the problem of identifying open quantum systems in the same setting that we do. The solution constructed in that work is related to the one that we will present below in the
sense that both consider observable time traces in the Laplace domain
and attempt to solve equations encoding the relation between the
unknown parameters and the measured signal. The principal difference
is that whereas Ref.  \cite{Schirmer:1452667} expresses the unknown
parameters directly in terms of the measured signal, we use model
realization theory to first construct a minimal model of the system
from the measurements, and then estimate the parameters from this
minimal model.
 
The remainder of this paper is organized as follows: in section
\ref{sec:prob} we formulate the system identification problem in the
Markovian open quantum systems context. Then in section
\ref{sec:id_alg} we describe the identification algorithm based on
model realization, and highlight some of the distinctions between this
algorithm and the corresponding algorithm for closed systems developed
in Ref. \cite{JunMohan:14}. We illustrate the open system algorithm
with an example in section \ref{sec:resonant}, and in addition, we present several case studies in Appendix \ref{sec:case_studies} that exemplify the types of equations that must be solved in order to identify the parameters. Finally, section \ref{sec:disc} concludes  the paper with a discussion of results and future directions.

\section{Problem formulation}
\label{sec:prob}
The problem of interest is the identification of unknown parameters
that dictate the dynamical behavior of a quantum system. The system
state is represented by a density matrix $\rho \in \C^{N\times N}$,
where $\rho\dg=\rho$, $\rho \geq 0$ ($\rho$ is a positive semidefinite matrix), and $\tr \rho=1$. We assume the
system dynamics is governed by a quantum Markovian master equation in
Lindblad form~\cite{Len-1987,Alicki:07}:
\begin{equation}
  \label{eq:9}
  \dot \rho= \CalL \rho \equiv \CalL_H\rho+\CalL_D\rho,
\end{equation}
where $\CalL_H$ is the generator of Hamiltonian dynamics
\begin{equation} 
  \label{eq:11}
   \CalL_H \rho=-i[H(\theta), \rho]
\end{equation}
with $\theta$ a vector of unknown parameters ($\hbar=1$), and $\CalL_D$
represents a general Lindblad dissipative generator
\begin{equation}
  \label{eq:10}
\CalL_D \rho=\frac12 \sum\nolimits_{j,k=1}^{N^2-1} g_{jk}
\left([F_j, \rho F_k^\dag]+[F_j\rho, F_k^\dag]\right).
\end{equation}
Here $\{i F_k\}$ is an orthonormal basis for the Lie algebra $\su(N)$,
and the Hilbert-Schmidt inner product is defined as $\langle iF_m,
iF_n \rangle \equiv \tr(F_m\dg F_n)$. Letting $\mG=\left(g_{jk}
\right)_{j,k=1}^{N^2-1}$, we have that $\mG\dg=\mG \ge 0$
\cite{Alicki:07}. Expanding the Hamiltonian in this orthonormal
basis, we obtain a parameterized Hamiltonian in the form:
\begin{equation}
  \label{eq:ham_theta}
H(\theta)= \sum\nolimits_{m=1}^M h_m(\theta) F_m,
\end{equation}
where $h_m=\tr(H F_m) \in \mathbb{R}$ are some known functions of
$\theta$. We take $h_m$ as the unknown parameters from now on since
solving $\theta$ from $h_m$ is an algebraic problem.  The unknown
parameters that we want to estimate are the Hamiltonian parameters
$h_m$ and the Lindblad coefficients $g_{jk}$.

First we define the structure constants $C_{jkl}\in \C$ of the Lie
algebra $\su(N)$ with respect to the orthonormal basis $\{i F_k\}$
through the commutator of basis elements:
\begin{equation}
  \label{eq:47}
[i F_j, i F_k]=\sum\nolimits_{l=1}^{N^2-1} C_{jkl} i F_l.
\end{equation}
For future use we also define the constants $D_{jkl}\in \C$ through the anticommutator of basis elements  \cite{Alicki:07}:
\begin{equation}
  \label{eq:48}
\{i F_j, i F_k\}=-\frac{2}{N}\delta_{jk} I +
\sum\nolimits_{l=1}^{N^2-1} D_{jkl} i F_l.  
\end{equation}
The dynamics of the expectation value of an observable
$F_n$, \ie $x_n(t)=\tr F_n \rho(t)$, can be written as
\begin{equation}
  \label{eq:15}
\dot x_n(t)=\tr{ F_n (\CalL_H \rho(t) + \CalL_D \rho(t)) }.
\end{equation}
From~\cite{Len-1987,JunMohan:14}, we derive that
\begin{equation}
  \label{eq:4}
\begin{aligned}
\dot x_n(t)=\sum\nolimits_{p=1}^{N^2-1} 
\left(Q_{np}+R_{np} \right)x_p(t)+b_n,   
\end{aligned}
\end{equation}
where
\begin{equation}
  \label{eq:41}
  \begin{aligned}
Q_{np}&=\sum\nolimits_{m=1}^{M}  C_{mnp} h_m,\\
R_{np}&= -  \frac14\sum\nolimits_{l,j,k=1}^{N^2-1} g_{jk}
\big(C_{njl} (C_{klp}+D_{klp}) \\
&\qquad\qquad  +C_{knl} (C_{ljp}+D_{ljp})\big),\\
b_n&=\frac{1}{N} \sum\nolimits_{j,k=1}^{N^2-1} \Im (g_{jk}) C_{njk}.
  \end{aligned}
\end{equation}
Collecting the $x_n$ in a vector $\mx \in \R^{N^2-1}$, we obtain a
linear differential equation describing the complete dynamics:
\begin{equation}
  \label{eq:2}
\dot{\mx}(t)= \mA \mx(t) +\mb, \quad x_n(0)= \tr F_n \rho(0),
\end{equation}
where the matrix $\mA \in \R^{(N^2-1) \times (N^2-1)}$ has elements
${A}_{np}=Q_{np}+R_{np}$, and the vector $\mb\in \R^{N^2-1}$
has elements $b_n$. 

The vector $\mx(t)$ is often called the \emph{coherence vector}, and is a complete representation of the
quantum state~\cite{Len-1987}. \erf{eq:2} gives an explicit
description of the quantum dynamics as a linear time invariant (LTI)
system, and hence it enables application of results from classical
linear systems theory. 

Next we turn to the observables being monitored. The measurement output consists of a vector of time-dependent observable expectation values:
\beq
\my(t) = \ma{\expect{O_1(t)}, & \expect{O_2(t)}, & \dots}^{\mathsf{T}}.
\eeq
In most physical systems one only has access to a limited set of observables (\eg local observables in a many-body system). However, as a result of the
dynamics, it is possible that all the parameters defining \erf{eq:9}
are imprinted in the time evolution of the monitored observables. This
is the notion we wish to formalize and exploit. We begin by expanding
each monitored observable $O_i$ in the basis $\{iF_n\}$ of the Lie
algebra $\su(N)$, \ie $O_i= \sum_{n=1}^{N^2-1} o^{(i)}_n F_n$.  With this expansion we can define the output vector as
\beq
\my(t) = \mC \mx(t)
\eeq
with the $i$-th row of the matrix $\mC$ being the elements $o^{(i)}_n$. Also  define the set $\mathcal{M} = \{F_{\bnu_1},
F_{\bnu_2}, ..., F_{\bnu_p}\}$, where $\bnu$ is a vector of length
$p$, as the collection of unique basis elements that appear in the expansion of all the measured observables.

We want to use the dynamical equation governing the time evolution of
only the observables being monitored to estimate the unknown parameters.  It is possible that the evolution does not couple the elements in $\mathcal{M}$ to all elements of the basis. This is equivalent to $\mA$ possessing block diagonalizable structure and the elements of $\mathcal{M}$ being coupled only through a proper subblock. A constructive procedure to find the relevant basis elements that couple to the measured observables is the following generalization of the filtration procedure in geometric control theory~\cite{Sastry:1999wd}. First we define the adjoint generator of dynamics through $\dot x_n = \tr(F_n (\mathcal{L} \rho)) = \tr((\mathcal{L}\dg F_n) \rho) 
$. Explicitly, $\CalL\dg X = \CalL_H\dg X + \CalL_D\dg X$, with
\bqa
\CalL_H\dg X = -i[X,H(\theta)] \nn 
\eqa
and
\bqa
\CalL_D\dg X =&\frac{1}{2}\sum_{j,k=1}^{N^2-1}g_{jk}
\left([F_j, F_k\dg X]+ [XF_j, F_k\dg]\right). \nn
\eqa
Then for the Lindblad dynamics in \erf{eq:9} define an iterative procedure as\bqa
G_0 &=& \mathcal{M}, \nn \\
G_i &=& \mathcal{L}\dg[G_{i-1}]\cup G_{i-1}, 
\label{eq:filtration}
\eqa 
where
\beq
\mathcal{L}\dg [G_{i-1}] = \{ F_j : \tr(F_j\dg \mathcal{L}\dg g)\neq 0, ~\textrm{where}~ g \in G_{i-1}\} \nn.
\eeq
That is, at each iteration we compute the adjoint evolution of each of the elements of $G_{i-1}$, and if the result has nonzero inner product with a basis element not already in $G_{i-1}$, this basis element is added to $G_i$. Since the dynamical system is finite dimensional, this iteration will saturate at a
maximal set $\bar G$ after finite steps, and we refer to this set as
the \emph{accessible set}.  Now writing all the $x_k$ with $F_k \in
\bar G$ in a vector $\mx_a$ of dimension $K \leq N^2-1$, the dynamics
for this vector is given by
\bqa
\label{eq:dyn_acc}
 \frac{{\rm d}}{{\rm d}t}\mx_a &=& \tilde\mA \mx_a + \tilde \mb, \\
 \my(t) &=& \tilde{\mC} \mx_a(t),
\eqa
where $\tilde{\mA}$ is a $K\times K$ sub-matrix of $\mA$, $\tilde
\mb$ is a $K\times 1$ sub-vector of $\mb$, and $\tilde{\mC}$ is a $p\times K$ sub-matrix of $\mC$; \ie all contain only the elements necessary to describe the evolution of the subset of observable averages collected in $\mx_a$, and how these define the measurement traces. 

Finally, we assume that the system is prepared in a fixed,
known initial state $\mx(0)$, and the corresponding initial state for
Eq.~\eqref{eq:dyn_acc} is $\mx_a(0)$. 

\section{Identification algorithm}
\label{sec:id_alg}
Suppose that we can measure the expectation values of some observables
in $\mathcal{M}$ at regular time instants $j\Delta t$, where $\Delta
t$ is the sampling period. Denote the measured expectation values of the observables as
$\{\my(j\Delta t)\}$ \footnote{Note that these expectation values may have to be collected from averaging
measurements on several runs of the experiment under the same initial
state.}, which is the output of the following
discretized form of Eq.~\eqref{eq:dyn_acc}:
\begin{equation}
\label{eq:dyn_disc}
\begin{aligned}
    \mx_a(j+1) =& \tilde\mA_d \left[\mx_a(j) + \int_{j\Delta t}^{(j+1)\Delta t} d\tau e^{-\tilde\mA \tau} \tilde\mb \right], \\
\my(j) = & \tilde\mC \mx_a(j),  
\end{aligned}
\end{equation}
where $\tilde\mA_d=e^{\tilde
  \mA \Delta t}$, and for brevity of notation we use $\mx_a(j) \equiv \mx_a(j \Delta
t)$ and $\my(j) \equiv \my(j \Delta t)$. 

\erf{eq:dyn_disc} defines a discrete time LTI system and we will now use invariants of different realizations of an LTI system to identify the unknown Hamiltonian and Lindblad parameters that generate the dynamics. To this end, we need to construct a realization from the measurement time traces.  There are many methods for constructing a realization of a linear dynamical system from input-output data in linear systems
theory~\cite{Callier:1991eo}. In~\cite{JunMohan:14}, we presented and utilized a
method called the eigenstate realization algorithm
(ERA)~\cite{Juang:1985kw} for identifying Hamiltonian dynamics, and in
the following we show that this method can be used in this case of
open system dynamics in Lindblad form also. For completeness, we
include the specification of ERA here.

\subsection{Eigenstate realization algorithm}
The first stage of the estimation algorithm is to construct a minimal
realization of the system based on input/output information. This is
achieved by ERA in three steps as follows.

\textbf{\underline{Step}} 1: \setlength{\leftmargin}{-10pt} Collect
the measured data into an $rp \times s$ matrix (generalized Hankel
matrix) as:
\begin{equation*}
  \label{eq:hankel}
  \begin{aligned}
 & \mH(k)=\\
&\ma{\my(k)&\my(k+t_1)&\cdots&\my(k+t_{s-1}) \\
\my(j_1+k)&\my(j_1+k+t_1)&\cdots&\my(j_1+k+t_{s-1}) \\
\vdots& \vdots &&\vdots \\
\my(j_{r-1}+k)&\my(j_{r-1}+k+t_1)&\cdots&\my(j_{r-1}+k+t_{s-1})}
  \end{aligned}
\end{equation*}
with arbitrary integers $j_i$ ($i=1$, $\cdots$, $r-1$) and $t_l$
($l=1$, $\cdots$, $s-1$). 

\textbf{\underline{Step}} 2: 
Find the singular value decomposition (SVD) of $\mH(0)$ as
\begin{equation}
  \label{eq:hankel_svd}
  \mH(0)=P\ma{\Sigma&0\\ 0 &0}Q\trp
=\ma{P_1&P_2}\ma{\Sigma&0\\ 0 &0}\ma{Q_1\trp\\ Q_2\trp}, \nn
\end{equation}
where $P\in \R^{rp\times rp}$, $Q\in \R^{s\times s}$ are both
orthonormal, and $\Sigma$ is a diagonal matrix with the non-zero
singular values of $\mH(0)$ determined up to numerical accuracy
$\epsilon$, \ie $\Sigma_{ii}>\epsilon$ for all $i\leq n_\Sigma$ where
$n_\Sigma$ is the dimension of $\Sigma$. The matrices $P_1$, $P_2$,
$Q_1$, $Q_2$ are partitions with compatible dimensions. When the measurement time traces are noisy, determining the cutoff parameter $\epsilon$ (and hence $n_\Sigma$) can be difficult since the noise can lead to non-decaying singular values. In this case, one can choose $n_\Sigma$ by demanding that the realization produced by ERA is of the correct order. We will return to this issue when we specify how the ERA realization will be used (see discussion after \erf{eq:realization_gen}).

\textbf{\underline{Step}} 3: Form a realization of the
system~\eqref{eq:dyn_disc} as $\hat \mA_d = \Sigma^{-\frac12} P_1\trp
\mH(1)Q_1\Sigma^{-\frac12}$, $\hat \mC = \mathsf{E}_{p}\trp P_1
\Sigma^{\frac12}$, and $\hat \mx(0) \equiv \Sigma^{\frac12} Q_1\trp e_1$, where $\mathsf{E}_p\trp = \left[ \mathbf{I}_p, 0_p,
  \cdots, 0_p\right]$ and $e_1$ is the first column of $\mathbf{I}_{s}$.  The triple $(\hat \mA_d, \hat \mC,  \hat \mx(0))$ forms a
realization of \erf{eq:dyn_disc}, and since the output data is an invariant of realizations, this triple generates the same output as the original system:
\begin{eqnarray} 
\my(j) = \hat \mC \hat \mA_d^j \hat \mx(0), \quad \text{for all } j\ge 0.
\label{eq:realization}
\end{eqnarray}  
Note that although the original system response is composed of a response to a non-zero initial state and a response to the forcing vector $\tilde\mb$ (\ie \erf{eq:dyn_disc}), the ERA realization is composed of only an initial state response. This is because at the level of input-output realizations it is not possible to distinguish between the response to initial states and the response to forcing inputs, and therefore ERA lumps all responses into one type.

This completes the specification of the ERA algorithm. It results in a
realization of the discrete time dynamical system~\eqref{eq:dyn_disc}
in the form of the triple $(\hat \mA_d, \hat \mC, \hat \mx(0))$.

We note that the measurements results do not have to be uniform in Step 1 of the algorithm; in particular, if it is known that some measurement results are particularly noisy or corrupted, those data points can be discarded by choosing appropriate integers $j_i$ ($i=1$, $\cdots$, $r-1$) and $t_l$ ($l=1$, $\cdots$, $s-1$). This data filtering can reduce the estimation error caused by measurement noise and outliers.

\subsection{Estimation algorithm}
\label{sec:estimation}
In order to estimate the parameters in the original dynamical system
we now convert the discrete-time LTI system realization obtained from
ERA, $(\hat \mA_d, \hat \mC, \hat \mx(0))$, to a continuous time
realization, by letting $\hat \mA \equiv \log \hat \mA_d/\Delta t$,
where the logarithm in this definition is the principal branch of the
natural logarithm. The accurate conversion from a discrete time system
to a continuous time system relies on the sampling time $\Delta t$
being sufficiently small to capture all continuous time dynamics.
For the Hamiltonian system identification case we were able to
  specify conditions on the sampling time based on the Shannon-Nyquist
  criteria, since in this case the output time traces $\my(t)$ are
  guaranteed to be band-limited \cite{JunMohan:14}. However, there is no such guarantee in the
Markovian open system case since now the time traces are generally
decaying oscillations, or \emph{time-limited} sigmals.  Therefore, we
do not currently have conditions on the sampling time for this
conversion to accurately provide a continuous-time realization (\ie to
avoid aliasing effects), but note that one can generally estimate a
valid sampling time from knowledge of the intrinsic frequencies in the
system. In engineering, it is typical practice to sample such
time-limited signals at $6$ or $8$ times the fastest frequency in the
signal \cite{Franklin:1997ua}, and such a heuristic suffices in our setting as well, as the
example in section \ref{sec:resonant} and the case studies in Appendix
\ref{sec:case_studies} bear out. Additionally, we note that although the Shannon-Nyquist criteria enables a formal specification of the minimum sampling time in the Hamiltonian estimation scenario, to apply it requires some knowledge of the system's spectrum. In the absence of this knowledge, the situation is the same in the closed and open systems cases: one needs to estimate a valid sampling time and possibly also try multiple sampling times.

Now, to estimate the unknown parameters, we use the fact that the system input-output relations are invariants of different realizations. We work in the Laplace domain in the following in order to form algebraic equations for the unknown
parameters. By equating the Laplace transform of the outputs of the original system and the ERA realization we get:
\beq
\label{eq:realization_gen}
\tilde{\mC} (s \mI-\tilde \mA)^{-1}\left(\mx_a(0) + \frac{\tilde\mb}{s}\right) 
=\hat \mC (s \mI-\hat \mA)^{-1}\hat\mx(0).
\eeq
This equation relates the unknown parameters to the measured data through the ERA realization. Explicitly, the right hand side of \erf{eq:realization_gen} is completely
determined by the measured data, and the left hand side is in terms of the
Hamiltonian parameters $h_m$ and Lindblad coefficients $g_{jk}$. The resolvent expressions on the right and left hand sides of \erf{eq:realization_gen}, $(s \mI-\tilde \mA)^{-1}$ and $(s \mI-\hat \mA)^{-1}$, can be computed symbolically, or alternatively the expressions can be expanded in powers of $s$, and the coefficients in this expansion can be equated to yield polynomial equations for the unknown parameters. Solving these multivariate polynomial equations leads to the identification of $g_m$ and $a_{jk}$. In the case studies encountered in Appendix~\ref{sec:case_studies} we are able to express the resolvents exactly by computing the matrix inverses symbolically, but one may need to resort to the expansion in $s$ in more complicated cases. We discuss this issue further in Section \ref{sec:disc}.

Note that $n_\Sigma$ in Step 2 of ERA dictates the maximum order of the polynomial on the right hand side of \erf{eq:realization_gen}. This suggests that we should choose $n_\Sigma$ to be the order of the denominator polynomial in the left hand side of \erf{eq:realization_gen}, which can be calculated from symbolic computations and obtained as an irreducible rational function. This choice coincides with the rank of $\mH(0)$ when there is no noise in the measurement time traces.

We highlight a crucial difference here between parameter estimation in the closed system and Markovian open system scenarios. For the former, $\mb=0$, and if the triple $(\tilde \mA, \tilde\mC, \mx_a(0))$ form a minimal realization (\ie is controllable and observable), then the Laplace transform on the left hand side of \erf{eq:realization_gen} is gauranteed to have a canonical form as a ratio of polynomials $Q(s)/P(s)$ \cite{Callier:1991eo}, with
\begin{equation}
  \label{eq:49}
  \begin{aligned}
P(s)&=\det(s \mI-\tilde\mA), \\
Q(s)&=\det \left( s\ma{\mI& \zero\\
\zero&\zero} - \ma{\tilde\mA& \mx_a(0)\\ \tilde\mC&\zero}\right).     
  \end{aligned}
\end{equation}
Having this form enabled us in Ref. \cite{JunMohan:14} to avoid explicitly computing the inverse $(s\mI - \tilde \mA)^{-1}$ or its expansion in a power series. These are both computationally intensive to compute since the matrix $\tilde \mA$ is a symbolic matrix containing the unknown parameters. However, in the open system case, where the left hand side of \erf{eq:realization_gen} has no known canonical form, one cannot avoid performing this symbolic inverse or power series computation. This is a critical difference in computational difficulty between the closed and open system parameter identification problems. 
 
We note that by converting back into the continuous time domain and formulating the Laplace transform relation in \erf{eq:realization_gen}, we have converted the problem of estimating parameters to one of solving polynomial equations. This is in contrast to directly estimating parameters using \erf{eq:realization}, which would involve solving transcendental equations for the unknown parameters. This simplification of the equations relating the unknown parameters to the measured data is one of the primary advantages of our approach. 
 
We conclude with two further comments on the above estimation algorithm:
\begin{enumerate}
\item The initial state may have to be chosen carefully. For example, if
$\mx_a$ is zero or an eigenvector of $\tilde \mA$, it leads to no
sensitivity in the output to any of the unknown parameters. Care must
be taken to avoid such degenerate cases. In fact, running the
algorithm with multiple initial states leads to more polynomial
equations of low order and thus can help to solve for the unknown parameters more
efficiently.
\item This system identification algorithm can result in multiple
  estimates of the unknown parameters, all of which satisfy the
  input/output relations captured by \erf{eq:realization_gen}. This is
  because several Markovian generators can generate the same map
  between an input state and measurement time trace, and hence are
  equivalent from an input/output perspective \cite{Burgarth:2012in}.
  When the algorithm results in multiple parameter estimates, one has
  to appeal to prior information, or add resources such as additional
  input states or observable time traces in order to uniquely specify
  a parameter set.
\end{enumerate}

The following section explicitly demonstrates the algorithm developed above and Appendix \ref{sec:case_studies} presents additional case studies.

\section{Dissipative energy transfer}
\label{sec:resonant}
In this section we apply the estimation algorithm developed above to a physically relevant example: energy transfer between two qubits at finite temperature. The qubits could represent the relevant energy levels of an atomic or molecular system, or spin-$\frac{1}{2}$ systems.
The Markovian master equation for the dynamics takes the form
\begin{eqnarray}
  \label{eq:64}
\frac{\mathrm{d} \rho}{\mathrm{d} t} &=& -i[H, \rho]  + \sum_{k=1}^2 \frac{\gamma_k}{2}(\sigma_z^k\rho\sigma_z^k-\rho) \nn \\
&& + \sum_{k=1}^2 2 g_k^- \left(\sigma_-^k \rho \sigma_+^k
-\frac12 \sigma_+^k \sigma_-^k \rho -\frac12
  \rho \sigma_+^k \sigma_-^k\right) \nn \\
&& ~ + 2 g_k^+ \left(\sigma_+^k \rho \sigma_-^k -\frac12 \sigma_-^k \sigma_+^k
 \rho -\frac12 \rho \sigma_-^k \sigma_+^k\right), 
\end{eqnarray}
where the Hamiltonian for the system is given by
\begin{equation}
  H= \frac{\omega_1}{2} \sigma_z^1 + \frac{\omega_2}{2} \sigma_z^2 + \delta_1
  \left(\sigma_+^1\sigma_-^{2} + \sigma_-^1 \sigma_+^{2} \right).
\end{equation}
This is a model for two possibly detuned qubits with coherent energy transfer dynamics, independent dephasing, and incoherent excitation-relaxation. The rates of excitation/relaxation ($g_k^+/g_k^-$) are functions of the temperature of the environment of the qubits \cite{Bre.Pet-2002}. There are nine parameters that dictate the dynamics in this model: $\theta = (\omega_1, \omega_2, \delta_1, \gamma_1, \gamma_2, g_1^+, g_1^-, g_2^+, g_2^-)$. For convenience, we define
\begin{equation}
  \label{eq:61}
  \begin{aligned}
\nu_1={g_1^+ +g_1^-}, \quad \mu_1={g_1^+ - g_1^-},\\
\nu_2={g_2^+ +g_2^-}, \quad \mu_2={g_2^+ - g_2^-}.
  \end{aligned}
\end{equation}

Assume that the observable being measured is $\bar z_1(t) \equiv\expect{\sigma_z^1(t)}$, or $\bar z_2(t) \equiv \expect{\sigma_z^2(t)}$, or both, since we will be interested in exploring the benefits of measuring multiple observables.
The state vector determined by the accessible set is the same
regardless of whether one or both of those observables are measured,
and is explicitly specified by
$\mx_a=[\la\sigma_z^1\ra$, $\la\sigma_z^2 \ra$, $\la
\sigma_x^1\sigma_x^2\ra$, $\la\sigma_x^1 \sigma_y^2 \ra$,
$\la\sigma_y^1\sigma_x^2 \ra$, $\la \sigma_y^1\sigma_y^2\ra]$, whereas
the dynamics of $\mx_a$ is determined by \erf{eq:dyn_acc} with
\begin{equation*}
  \label{eq:36}
  \begin{aligned}
&\quad \tilde \mA=\\
&\ma{ -2 \nu_1& 0& 0& \delta_1& -\delta_1& 0\\
 0& -2 \nu_2& 0& -\delta_1& \delta_1& 0\\
 0& 0& -\nu_s-\gamma_s& \omega_2& \omega_1& 0\\
 -\delta_1& \delta_1& -\omega_2& -\nu_s-\gamma_s& 0& \omega_1\\
 \delta_1& -\delta_1& -\omega_1& 0& -\nu_s-\gamma_s& \omega_2\\
 0& 0& 0& -\omega_1& -\omega_2& -\nu_s-\gamma_s},\\
&\quad \tilde\mb=\ma{\mu_1,&\mu_2,&0,&0,&0,&0}^\mathsf{T},
  \end{aligned}
\end{equation*}
where $\nu_s=\nu_1+\nu_2$ and $\gamma_s=\gamma_1+\gamma_2$.  
The matrix defining the output depends on which observable is being
monitored and is given by
\begin{equation}
  \label{eq:23}
  \begin{aligned}
\expect{\sigma_z^1}: &\quad \tilde\mC_{z_1}=\ma{1,0,0,0,0,0}, &\text{or} \\
\expect{\sigma_z^2}: &\quad \tilde\mC_{z_2}=\ma{0,1,0,0,0,0}, &
  \end{aligned}
\end{equation}
or the concatenation of these two row vectors into a $2\times 6$ matrix, $\tilde\mC_{z_1z_2}$ if both observables are being monitored.

Letting
the initial state be
$\ket{\psi_1(0)}=\ket{0} \otimes (\ket{0}+\ket{1})/\sqrt{2}$,
we have  that
 \begin{equation}
   \label{eq:65}
 \mx_{a}(0)=\ma{1,&0,&0,&0,&0,&0}^\mathsf{T}.
 \end{equation}
 Since the matrix $\tilde \mA$ is small, we can symbolically calculate the resolvents in \erf{eq:realization_gen}, to obtain the Laplace transform of the two possible measurement traces $\bar z_1(t)$ or $\bar z_2(t)$:
 \begin{equation}
   \label{eq:res_trans_Z1}
  \begin{aligned}
\bar{Z}_1(s)&=\tilde\mC_{z_1}(s\mI-\tilde\mA)^{-1}\mx_a(0)+\tilde\mC_{z_1}(s\mI-\tilde\mA)^{-1}\tilde\mb/s\\
&=\frac{s^4 + q_3 s^3+q_2s^2+q_1s+q_0}{s^5+p_4s^4+p_3s^3+p_2s^2+p_1s}    
  \end{aligned}
\end{equation}
with
\begin{eqnarray}
  \label{eq:34}
q_3 &=& \mu_1 + 2\nu_1 + 4\nu_2 + 2\gamma_s, \nn \\
q_2 &=& 2\delta_1^2 + \nu_1^2 + 6\nu_1\nu_2 + 2\nu_1\gamma_s + 2\mu_1\nu_1 + 5\nu_2^2 + 6\nu_2\gamma_s \nn \\
&& + 4\mu_1\nu_2 + \gamma_s^2 + 2\mu_1\gamma_s + \omega_d^2, \nn \\
q_1 &=& 2\delta_1^2\mu_1 + 2\delta_1^2\mu_2 + 2\delta_1^2\nu_1 + 2\delta_1^2\nu_2 + 2\delta_1^2\gamma_s + \mu_1\nu_1^2  \nn \\
&& + 5\mu_1\nu_2^2 + 4\nu_1\nu_2^2 + 2\nu_1^2\nu_2 + \mu_1\gamma_s^2 + 2\nu_2\gamma_s^2 + 4\nu_2^2\gamma_s \nn \\
&& + \mu_1\omega_d^2 + 2\nu_2\omega_d^2 + 2\nu_2^3 + 6\mu_1\nu_1\nu_2 + 2\mu_1\nu_1\gamma_s \nn \\
&& + 6\mu_1\nu_2\gamma_s + 4\nu_1\nu_2\gamma_s \nn \\
q_0 &=& 2\mu_1\nu_2^3 + 2\delta_1^2\mu_1\gamma_s + 2\delta_1^2\mu_2\gamma_s + 4\mu_1\nu_1\nu_2^2 + 2\mu_1\nu_1^2\nu_2 \nn \\
&& + 2\mu_1\nu_2\gamma_s^2 + 4\mu_1\nu_2^2\gamma_s + 2\mu_1\nu_2\omega_d^2 + 2\delta_1^2\mu_1\nu_1 \nn \\
&& + 2\delta_1^2\mu_1\nu_2 + 2\delta_1^2\mu_2\nu_1 +
2\delta_1^2\mu_2\nu_2 + 4\mu_1\nu_1\nu_2\gamma_s, 
\end{eqnarray}
and
\begin{eqnarray}
  \label{eq:66}
p_4 &=& 4\nu_1 + 4\nu_2 + 2\gamma_s, \nn \\
p_3 &=& 4\delta_1^2 + 5\nu_1^2 + 14\nu_1\nu_2 + 6\nu_1\gamma_s + 5\nu_2^2 + 6\nu_2\gamma_s \nn \\
&& + \gamma_s^2 + \omega_d^2, \nn \\
p_2 &=& 8\delta_1^2\nu_1 + 8\delta_1^2\nu_2 + 4\delta_1^2\gamma_s + 2\nu_1^3 + 14\nu_1^2\nu_2 + 4\nu_1^2\gamma_s \nn \\
&& + 14\nu_1\nu_2^2 + 16\nu_1\nu_2\gamma_s + 2\nu_1\gamma_s^2 + 2\nu_1\omega_d^2 + 2\nu_2^3 \nn \\
&& + 4\nu_2^2\gamma_s + 2\nu_2\gamma_s^2 + 2\nu_2\omega_d^2, \nn \\
p_1 &=& 4\delta_1^2\nu_1^2 + 8\delta_1^2\nu_1\nu_2 + 4\delta_1^2\nu_1\gamma_s + 4\delta_1^2\nu_2^2 + 4\delta_1^2\nu_2\gamma_s \nn \\
&& + 4\nu_1^3\nu_2 + 8\nu_1^2\nu_2^2 + 8\nu_1^2\nu_2\gamma_s + 4\nu_1\nu_2^3 + 8\nu_1\nu_2^2\gamma_s \nn \\
&& + 4\nu_1\nu_2\gamma_s^2 + 4\nu_1\nu_2\omega_d^2, 
\end{eqnarray}
where $\omega_d=\omega_1-\omega_2$. And,
\begin{equation}
   \label{eq:res_trans_Z2}
  \begin{aligned}
\bar{Z}_2(s)&=\tilde\mC_{z_2}(s\mI-\tilde\mA)^{-1}\mx_a(0)+\tilde\mC_{z_2}(s\mI-\tilde\mA)^{-1}\tilde\mb/s\\
&=\frac{r_3 s^3+r_2s^2+r_1s+r_0}{s^5+p_4s^4+p_3s^3+p_2s^2+p_1s} ,   
  \end{aligned}
\end{equation}
with
\begin{eqnarray}
  \label{eq:67}
r_3 &=& \mu_2, \nn \\
r_2 &=& 2\delta_1^2 + 4\mu_2\nu_1 + 2\mu_2\nu_2 + 2\mu_2\gamma_s, \nn \\
r_1 &=& 2\delta_1^2\mu_1 + 2\delta_1^2\mu_2 + 2\delta_1^2\nu_1 + 2\delta_1^2\nu_2 + 2\delta_1^2\gamma_s + 5\mu_2\nu_1^2 \nn \\
&& + \mu_2\nu_2^2 + \mu_2\gamma_s^2 + \mu_2\omega_d^2 + 6\mu_2\nu_1\nu_2 + 6\mu_2\nu_1\gamma_s \nn \\
&& + 2\mu_2\nu_2\gamma_s, \nn \\
r_0 &=& 2\mu_2\nu_1^3 + 2\delta_1^2\mu_1\gamma_s + 2\delta_1^2\mu_2\gamma_s + 2\mu_2\nu_1\nu_2^2 + 4\mu_2\nu_1^2\nu_2 \nn \\
&& + 2\mu_2\nu_1\gamma_s^2 + 4\mu_2\nu_1^2\gamma_s + 2\mu_2\nu_1\omega_d^2 + 2\delta_1^2\mu_1\nu_1 \nn \\
&& + 2\delta_1^2\mu_1\nu_2 + 2\delta_1^2\mu_2\nu_1 +
2\delta_1^2\mu_2\nu_2 + 4\mu_2\nu_1\nu_2\gamma_s 
\end{eqnarray}
and $p_i$ defined as in \erf{eq:66}.

It is clear from these expressions that we can identify $\nu_i$,
$\mu_i$, which then allows identification of $g_i^+$, $g_i^-$. At the
same time, note that only the linear combinations $\gamma_s =
\gamma_1+\gamma_2$ and $\omega_d = \omega_1-\omega_2$ occur in the
above equations but not the individual parameters $\gamma_i$ and
$\omega_i$ ($i=1,2$). This implies that only these linear combinations
can be determined from the measurements and the initial state, but not
the individual parameters that enter them. These linear combinations
describe the energy difference between the qubits and the
dephasing-induced broadening of this energy difference. The physical
observables encoding average population of the excited state of either
qubit only allow determination of these collective (in the case of
$\gamma_s$) or relative (in the case of $\omega_d$) properties of the
system. Furthermore, another restriction that we can immediately
observe is that only even powers of $\delta_1$ and $\omega_d$ occur in
all of the polynomials above, and therefore we only expect to
determine these parameters up to a sign difference. The signs of these
parameters are not estimable by the local measurements we have chosen,
and prior knowledge, or additional initial states or measurements are
necessary to identify these signs.

\begin{figure}[bt]
  \includegraphics[scale=0.48]{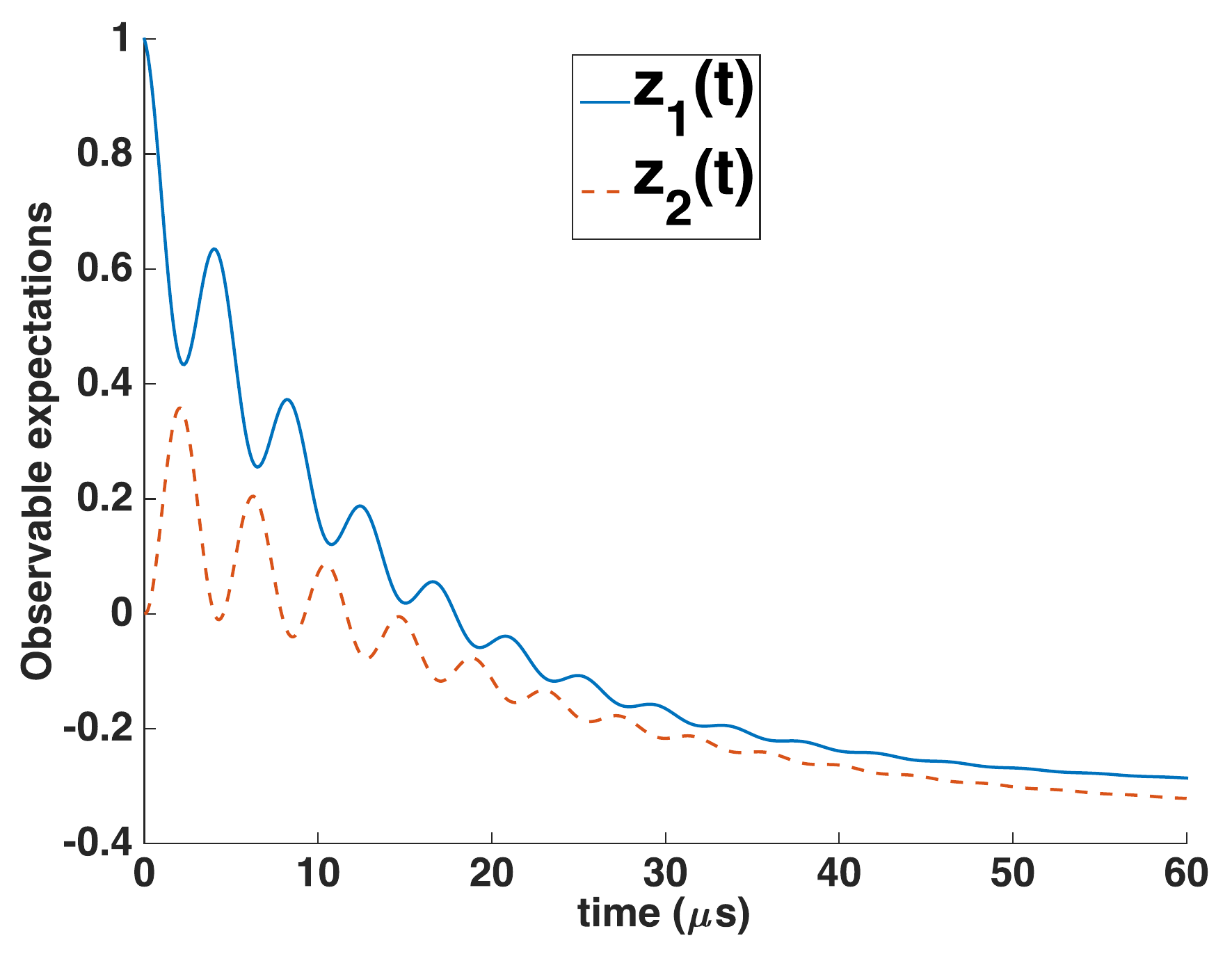}
  \caption{(Color online) Measurement time traces of $\expect{\sigma_z^1(t)}$ and $\expect{\sigma_z^2(t)}$ for the energy transfer example presented in section \ref{sec:resonant}. The initial state and system parameters are specified in the main text. \label{fig:traces_noiseless}}
\end{figure}

Before we present the results of the estimation simulation, we outline four possible modes of estimation that could be performed given that there are two possible observables: 
\begin{enumerate}
\item In \texttt{mode 1}, only the measurement trace $\bar{z}_1(t)$ is used to construct the Hankel matrix and system realization ($\tilde\mC = \tilde\mC_{z_1}$), and the polynomial system used to solve for the unknown parameters is constructed using only $\bar{Z}_1(s)$.
\item In \texttt{mode 2}, only the measurement trace $\bar{z}_2(t)$ is used to construct the Hankel matrix and system realization ($\tilde\mC = \tilde\mC_{z_2}$), and the polynomial system used to solve for the unknown parameters is constructed using only $\bar{Z}_2(s)$.
\item In \texttt{mode 3}, the measurement traces $\bar{z}_1(t)$ \emph{and} $\bar{z}_2(t)$ are used to construct the Hankel matrix and system realization ($\tilde\mC = \tilde\mC_{z_1z_2}$), and the polynomial system used to solve for the unknown parameters is constructed using only $\bar{Z}_1(s)$.
\item In \texttt{mode 4}, the measurement traces $\bar{z}_1(t)$ \emph{and} $\bar{z}_2(t)$ are used to construct the Hankel matrix and system realization ($\tilde\mC = \tilde\mC_{z_1z_2}$), and the polynomial system used to solve for the unknown parameters is constructed using only  $\bar{Z}_2(s)$.
\end{enumerate}
Obviously, \texttt{modes 1} and \texttt{2} are practically more
  attractive since they only involve collecting one observable's time
  trace. However, there may be a benefit to consider  \texttt{modes
  3} and \texttt{4} since in these modes of estimation one is
  providing the algorithm with more data with which to construct the
  realization (although the accessible vector does not change, and so some of this additional data is redundant). One could also imagine a fifth mode where one uses the measurement traces $\bar{z}_1(t)$ \emph{and} $\bar{z}_2(t)$ to construct the Hankel matrix and system realization ($\tilde\mC = \tilde\mC_{z_1z_2}$), and then constructs the polynomial system from the definitions of \emph{both} $\bar{Z}_1(s)$ and $\bar{Z}_2(s)$, \ie some polynomial equations from the definition of the coefficients in $\bar{Z}_1(s)$ and some from the definition of the coefficients in $\bar{Z}_2(s)$. For our example, where the number of parameters is small enough such that one can get enough equations from using the definition of just one
  of the Laplace transforms, we did not find any advantage to using
  this fifth mode of estimation, and so do not investigate it further. 

To illustrate the algorithm we fix the nominal parameters of the system
as $\omega_1=1.3$MHz, $\omega_2=2.4$MHz, $\delta_1=0.5$MHz,
$\gamma_1=0.03$ MHz, $\gamma_2=0.035$MHz, $g^+_i =
0.02\bar{n}(\omega_i)$MHz, $g^-_i = 0.02(\bar{n}(\omega_i)+1)$MHz,
with 
\beq 
\bar{n}(\omega) = \frac{1}{e^{\frac{\omega}{k_B T}}-1} \nn
\eeq 
being the Bose-Einstein distribution at temperature $T$. In the following we fix the temperature at $k_B T =
0.8\omega_1$. For a system with these parameters, we generate a time
trace of $\bar{z}_1(t)$ and $\bar{z}_2(t)$ from $t=0$ to $t=t_{\rm f}\equiv
60\mu$s, with $\Delta t=0.01 \mu$s. The resulting time traces are shown
in Fig. \ref{fig:traces_noiseless}. We form the Hankel matrix using
either one or both of these time traces, depending on the mode of estimation, with $t_i=j_i=1$ (\ie using every data point), and $r=s=\frac{t_{\rm f}/\Delta t}{2}$. The ERA
realization of the discrete time system is formed with $n_\Sigma =
\text{rank}(\mH(0))$, and the corresponding continuous time system is
formed using the prescription given in section \ref{sec:estimation}.
Finally, the Laplace transform expression on the right-hand-side of
\erf{eq:realization_gen} is computed to obtain \footnote{We used the
  Matlab Control System Toolbox function \texttt{tf} in order to
  obtain these Laplace transforms.}
 \begin{equation}
  \begin{aligned}
\bar{Z}^{est}_1(s)&=\frac{s^4 + 0.2702s^3+1.7302s^2+0.072s-0.0034}
{s^5+0.3624s^4+2.2569s^3+0.3243s^2+0.011s} \nn
  \end{aligned}
\end{equation}
and
 \begin{equation}
  \begin{aligned}
\bar{Z}^{est}_2(s)&=\frac{-0.0176 s^3+0.4944s^2+0.0209s-0.0039}
{s^5+0.3624s^4+2.2569s^3+0.3243s^2+0.011s}. \nn    
  \end{aligned}
\end{equation}
Combining these expressions with corresponding Laplace transforms
Eqs.~\eqref{eq:res_trans_Z1} and \eqref{eq:res_trans_Z2}, we obtain a system of
polynomial equations for the unknown parameters (we can choose the
seven simplest equations since there are seven unknown parameters). We
solved these equations using \texttt{PHClab}, the Matlab interface to
the \texttt{PHCPack} libraries for solving polynomial systems
\cite{Verschelde:08}. 

\begin{table*}[t]
\centering
\begin{tabular}{|l|c|c|c|c|c|c|c|}
\hline
\textbf{Parameters} & $\omega_d$ & $\delta_1$ & $\nu_1$ & $\nu_2$ & $\mu_1$ & $\mu_2$ & $\gamma_s$ \\ \hline
Nominal values & \textbf{-1.1} & \textbf{0.5} & \textbf{0.0361} & \textbf{0.022} & \textbf{-0.02} & \textbf{-0.0176} & \textbf{0.065} \\ \hline
\multirow{2}{*}{}\texttt{Mode 1} & $\pm 1.0973$ & $\pm 0.5029$ & 0.0677 & -0.0096 & 0.0432 & -0.0815 & 0.065 \\ \cline{2-8} 
                  & $\pm 1.1$ & $\pm 0.5$ & 0.0361 & 0.022 & -0.02 & -0.0176 & 0.065  \\ \hline
\texttt{Mode 2} & $\pm 1.1$ & $\pm 0.5$ & 0.0361 & 0.022 & -0.02 & -0.0176 & 0.065 \\ \hline
\end{tabular}
\caption{Estimates derived from noiseless measurement traces. The modes of estimation (\texttt{mode 1} and \texttt{mode 2}) are explained in the main text.  \label{tab:noiseless}}
\end{table*}
  
  Table \ref{tab:noiseless} shows the results of solving for the
  parameters when the measurements are noiseless. We found no
  difference between \texttt{modes 1} and \texttt{3}, and also between
  \texttt{modes 2} and \texttt{4}, when the measurements are
  noiseless, and therefore we only present results from \texttt{modes
    1} and \texttt{2} of estimation in table \ref{tab:noiseless}.  For
  \texttt{mode 1}, the first observation is that there exist several
  estimates that deviate from the nominal values as shown on the
  first line. This is because in this case we choose the lowest order
  polynomials defined by the coefficients of $\bar{Z}_1(s)$ and the
  resulting polynomial system has multiple solutions, among which the
  nominal set is only one of them. Secondly, as expected from the
  above observation that only even orders of $\delta_1$ and $\omega_d$
  enter the polynomial system, the signs of these parameters are
  indeterminate. In estimation \texttt{mode 2}, the estimate quality
  increases and in fact, the only uncertainty is in the sign of
  $\delta_1$ and $\omega_d$. This is an example of how the choice of
  observable dictates the quality of the estimation.

In summary, we see that given noiseless measurement records, the above estimation algorithm can determine the unknown Hmailtonian and Lindblad parameters to within the limitations of the data (\eg in the above example, the limitations were that $\omega_1$, $\omega_2$, $\gamma_1$, and $\gamma_2$ are not individually estimable and that the signs of $\delta_1$ and $\omega_d$ are not estimable). 

\section{Noisy measurements}
\label{sec:noisy}
In this section we investigate the robustness of the system
identification algorithm by reexamining the two-qubit dissipative energy transfer example with noisy measurement traces. In principle, the noise on expectation values of observables can be made arbitrary small since the signal-to-noise decreases as $1/\sqrt{N}$, where $N$ is the number of measurements that are averaged to estimate the expectation value of the observable.  Because of this, measurement noise is especially small in ensemble systems like NMR \cite{Hou:2014wg}; however, in systems without natural access to ensembles, \eg a single superconducting qubit, noise on time traces cannot be neglected in practical situations and we must assess the robustness of the above system identification algorithm to noise.

We consider the same system as in section \ref{sec:resonant}, with
access to time traces of one or both observables:
$\expect{\sigma_z^1(t)}$ and $\expect{\sigma_z^2(t)}$. Suppose that
these time traces are corrupted by additive Gaussian noise
\begin{equation*}
  z_i^{\text{noisy}}(j) = \bar{z}_i(j) + \xi_i(j),
\end{equation*}
where $\xi_i(j) \sim \mathcal{N}(0, \sigma^2)$ for all $i,j$. Since
the expectation values are formed by averaging many independent
measurement outcomes, this is a reasonable model for the noise by the
Central Limit Theorem. We construct the Hankel matrix and perform the
estimation in exactly the same way as section \ref{sec:resonant}, with
the only difference being that $n_\Sigma$ is fixed to be $5$ instead
of the rank of the Hankel matrix $\mH(0)$, since the order of the denominator polynomial on the left hand side of \erf{eq:realization_gen} (which takes the form in Eqs. (\ref{eq:res_trans_Z1}) or (\ref{eq:res_trans_Z2}) for this example) is $5$. 

To assess the quality of estimation, we compute the relative error in
estimation for each of the seven parameters as 
\begin{equation*}
e_i=\left|\frac{\hat{\theta}_i - \theta_i}{\theta_i}\right| \times 100 \%,
\end{equation*}
where $\theta_i$ and $\hat{\theta}_i$ are the nominal and estimated
values of the parameter, respectively. If the estimation produces
multiple solutions, then we choose the one with the least sum of
errors $\sum_{i=1}^7 e_i$. We generate $M=500$ instances of
noise with given standard deviation $\sigma$, and calculate the
estimation errors for each instance. These errors are then averaged to
yield a mean relative error $\bar{e}_i$, which captures the
performance of the algorithm. We evaluate these mean relative errors for standard
deviations $\sigma=0.05$, $0.10$, $0.15$, as well as for the noiseless case
($\sigma=0$). The initial state in all instances is the same as
in section \ref{sec:resonant}, and the only difference in parameters
is that the duration of the measurement time traces is longer: $t_{\rm f}=120
\mu$s.

The first observation from this simulation (data not shown) is that there is
a difference between \texttt{modes} \texttt{1} and \texttt{3} of estimation
(and \texttt{modes} \texttt{2} and \texttt{4}) when the measurements are noisy; it is beneficial to use data from both measurement time traces, $z_1^{\text{noisy}}(t)$ and
$z_2^{\text{noisy}}(t)$, to form the realization even if the polynomial system
is formed from the Laplace transform of one of the time traces. Therefore we  present only the  estimation \texttt{modes} \texttt{3} and \texttt{4}. 

\begin{figure}[bt]
  \includegraphics[scale=0.5]{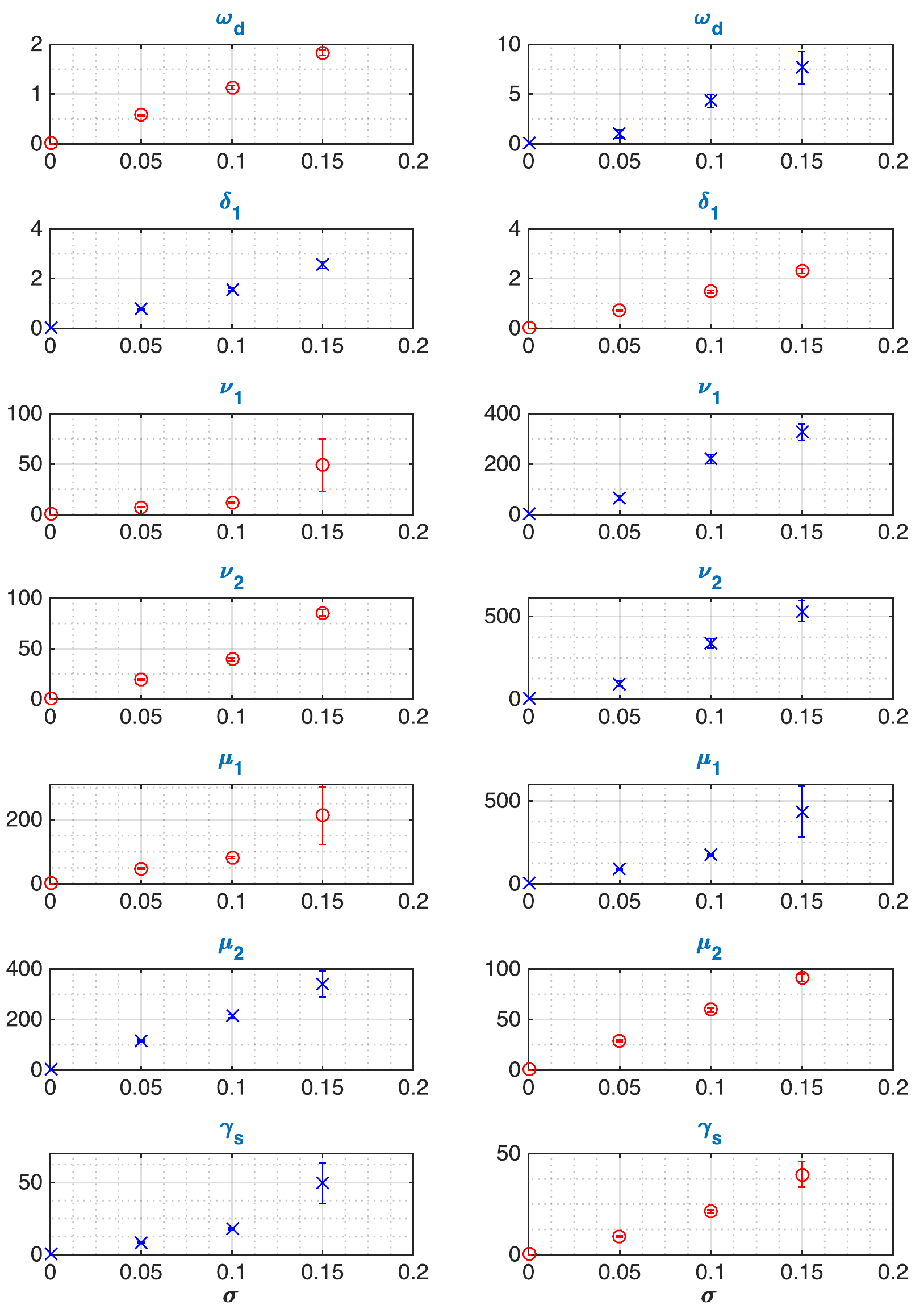}
  \caption{\label{fig:noisy_err} (Color online) Average relative error of parameter estimates as a function of the standard deviation ($\sigma$) of the additive noise on the measurement traces. 
 In all the plots, the $y$-axis shows the relative error as a percentage and the $x$-axis is $\sigma$. The left (right) column shows the error for estimation under \texttt{mode 3} (\texttt{mode 4}). The red circles denote the result of the better estimation mode for that parameter (lower maximum error), and the error bars indicate the standard error for the average (which is taken over $M=500$ instances of noise).}
\end{figure}

Figure \ref{fig:noisy_err} shows the mean relative error for each
parameter as a function of the standard deviation of the measurement
noise. The left (right) column plots $\bar e_i$ under \texttt{mode 3} (\texttt{mode 4}) of
estimation.
This figure shows that the average error in estimation is small for small $\sigma$, but quickly becomes quite large. An interesting
feature is that the performance can be very different under
\texttt{mode 3} and \texttt{mode 4} of estimation, with one performing
better for some parameters and worse for others. Also, the performance
can vary significantly among parameters. Estimates of the Hamiltonian parameters, $\omega_d$ and $\delta_1$, exhibit the greatest robustness to measurement noise, while estimates of the open system parameters are more sensitive. In particular, estimates of $\mu_1, \mu_2$ and $\nu_2$, which are the parameters of smallest magnitude, suffer the most from measurement noise. It is clear that this open system identification algorithm is not as robust as the Hamiltonian version \cite{JunMohan:14}. However, in the regions of low noise ($\sigma < 0.1$) relevant for NMR and other ensemble experimental platforms, the performance is acceptable, especially for the Hamiltonian parameters, which can be estimated well even in the presence of dissipation and dephasing. 

\section{Discussion}
\label{sec:disc}

We have extended the quantum system identification approach developed
in Ref. \cite{JunMohan:14} for Hamiltonian systems, to Markovian open
quantum systems. The approach proceeds by forming a realization of the
quantum system from (time-dependent) input-output data and then uses
this realization to form a system of polynomial equations for the
parameters that define the system. The strengths of the approach are
its ability to incorporate prior information and its ability to
produce parameter estimates even with time dependent measurements of
only a few observables.

As with the Hamiltonian case dealt with in Ref. \cite{JunMohan:14}, having access to time-dependent data enables us to directly estimate the generator of dynamics, which has a significant advantage in that it is specified by fewer parameters than the dynamical map at a fixed time (\eg Kraus map or unitary). In contrast to the Hamiltonian case, the Markovian open system identification scenario ideally requires the symbolic computation of a resolvent as expressed in \erf{eq:realization_gen}, which can be computationally expensive for large systems. Alternatively, as mentioned in Section \ref{sec:estimation} one could avoid the resolvent computation by using a power series expansion of the resolvent. This approach is equivalent to using the Markov parameters of an LTI system as the model realization invariant, as opposed to the Laplace transform of the output. Although this is computationally advantageous, we have found that the Markov parameters are more susceptible to measurement noise and therefore this approach is expected to yield less robust estimates of the parameters. As a consequence, we expect that symbolic computation of the resolvent will be the fundamental limitation to applying this approach to identification of large Markovian open quantum systems. However, as demonstrated in Sections \ref{sec:case_studies} and \ref{sec:resonant}, the required symbolic computation can be easily performed for small open systems, which are still difficult to identify using other approaches. 

The noise robustness results in \ref{sec:noisy} suggest that a direction for future work is to understand exactly why this realization-based system identification algorithm is more robust to measurement noise in the Hamiltonian (closed system) case than in the open system case. One clue as to why this may be is that Markovian open system parameters almost always dictate exponential rates of decay of measurement traces, and parameter fitting to decaying exponentials is a notoriously ill-conditioned problem \cite{Pereyra:2010wt}. More work is required to understand the properties of this algorithm under measurement noise and to increase its robustness in the open system scenario.

\begin{acknowledgments}
  This work was supported by the Laboratory Directed Research and
  Development program at Sandia National Laboratories. Sandia is a
  multi-program laboratory managed and operated by Sandia Corporation,
  a wholly owned subsidiary of Lockheed Martin Corporation, for the
  United States Department of Energy's National Nuclear Security
  Administration under contract DE-AC04-94AL85000.  JZ thanks the
  financial support from NSFC under Grant No.  61174086, State Key
  Laboratory of Precision Spectroscopy, ECNU, China, and
  Project-sponsored by SRF for ROCS SEM. 
\end{acknowledgments}

\pagebreak
\begin{widetext}
\appendix
\section{Case studies}
\label{sec:case_studies}
In this appendix we present some examples that illustrate the
Markovian open system identification algorithm developed in this
paper. The system studied in all examples is a 1D chain of $n$-qubits,
with closed system dynamics governed by the Hamiltonian
\begin{equation}
  \label{eq:ham_eg}
  H=\sum_{k=1}^n \frac{\omega_k}{2} \sigma_z^k+\sum_{k=1}^{n-1} \delta_k
  \left(\sigma_+^k\sigma_-^{k+1} + \sigma_-^k \sigma_+^{k+1} \right). 
\end{equation}
This Hamiltonian is often used as a model for a spin ``wire'' that
enables quantum state transfer \cite{Bose:2007ji}, and is a relevant to resonant energy transfer mechanisms between atomic or solid-state systems, \eg \cite{Ryabtsev:2010do}.

In Ref. \cite{JunMohan:14} we analyzed the identification of the Hamiltonian parameters of this closed system when the observable being directly measured is $\expect{\sigma_x^1}$. When the generalized Pauli operators is chosen as the basis of operators $\{F_k\}$, the accessible set was found to be $\bar{G} = \{ 2^{-n/2}\sigma_x^1, 2^{-n/2}\sigma_y^1 \} \cup \{2^{-n/2}\sigma_z^1 \cdots \sigma_z^{k-1}\sigma_x^k, 2^{-n/2}\sigma_z^1 \cdots \sigma_z^{k-1}\sigma_y^k \}_{k=2}^n$.
Then denoting the average of the observables in $\bar{G}$ as $\mx_a = \left[
  \bar x_1, \bar y_1, ..., \bar x_n, \bar y_n \right]^{\mathsf{T}}$, where $\bar
x_1=\expect{\sigma_x^1}, \bar y_1=\expect{\sigma_y^1}$ and $\bar x_k
=\expect{\sigma_z^1 \cdots \sigma_z^{k-1}\sigma_x^k}, \bar y_k
=\expect{\sigma_z^1 \cdots \sigma_z^{k-1}\sigma_y^k}$ for $k\ge
2$, the dynamics of $\mx_a$ is determined by
\begin{equation}
  \label{eq:26}
\dot \mx_a=\tilde \mA \mx_a,
\end{equation}
where the system matrix $\tilde \mA$ is given by~\cite{JunMohan:14}:
\begin{equation}
  \label{eq:22}
  \tilde \mA=\left[\begin{array}{cccccccc}
0&  \omega_1&   0& -\delta_1&   &   &   &   \\
-\omega_1& 0&  \delta_1&   0&  0 &   &   &   \\
0& -\delta_1& 0&  \omega_2&   0& \ddots&   &   \\
\delta_1&   0& -\omega_2& 0&  \ddots&  \ddots&   0&   \\
&   0&   \ddots& \ddots&   \ddots&  \ddots& \ddots& -\delta_{n-1}\\
&   &  \ddots&   \ddots& \ddots&  0&  \delta_{n-1}&   0\\
&   &   &   0&   0& -\delta_{n-1}& 0&  \omega_n\\
&   &   &   &  \delta_{n-1}&   0& -\omega_n& 0
 \end{array}\right].
\end{equation}
Finally, in this basis the direct observation of $\expect{\sigma_x^1}$ corresponds to $\mC=\left[ 1, 0, 0, ..., 0 \right]$. 

In the following we will consider various types of open system dynamics for this 1D spin chain and apply the algorithm developed above to estimate the parameters defining the Hamiltonian and the open system dynamics. 

\subsection{Independent dephasing}
\label{sec:ind_dephasing}
We first consider a common decoherence model for spin-$\frac12$ systems, namely, independent Markovian dephasing on each spin~\cite{Bre.Pet-2002}.  This is described by the Lindblad terms
\begin{equation}
  \label{eq:42}
  \sum_{k=1}^{n} 2 \gamma_k(\sigma_z^k\rho\sigma_z^k-\rho),
\end{equation}
where the superscript $k$ indexes the spin, and the factor 2 is for
the ease of expression. Comparing Eqs.~\eqref{eq:10}
and~\eqref{eq:42}, we see that the independent dephasing model is
equivalent to the choice
\begin{equation}
  \label{eq:43}
 g_{lk}=
\begin{cases}
  \gamma_k \delta_{lk}, & \text{if } \tr (X_k \sigma_z^k)\neq 0; \\
0, & \text{otherwise},
\end{cases}
\end{equation}
meaning that the Lindblad coefficient matrix $\mG$ is diagonal and real. Again, assuming that the directly measured quantity is $\expect{\sigma_x^1}$, the system matrix $\tilde \mA$ now becomes:
\begin{equation}
  \label{eq:44}
\tilde \mA=\left[\begin{array}{cccccccc}
-\gamma_1&  \omega_1&   0& -\delta_1&   &   &   &   \\
-\omega_1& -\gamma_1&  \delta_1&   0&  0 &   &   &   \\
0& -\delta_1&   -\gamma_2&  \omega_2&   0& \ddots&   &   \\
\delta_1&   0& -\omega_2&   -\gamma_2&  \ddots&  \ddots&   0&   \\
&   0&   \ddots& \ddots&   \ddots&  \ddots& \ddots& -\delta_{n-1}\\
&   &  \ddots&   \ddots& \ddots&   -\gamma_{n-1}&  \delta_{n-1}&   0\\
&   &   &   0&   0& -\delta_{n-1}&   -\gamma_n&  \omega_n\\
&   &   &   &  \delta_{n-1}&   0& -\omega_n&   -\gamma_n
 \end{array}\right].
\end{equation}
Comparing \Eqs{eq:22} and~\eqref{eq:44}, we observe that the
dephasing terms introduce nonzero diagonal elements in
$\tilde\mA$, whereas the accessible set $\bar{G}$ remains the same. In addition, for these dynamics, $\tilde \mb=0$. All parameters in the Hamiltonian \erf{eq:ham_eg} and Lindblad coefficients in \erf{eq:42} appear in $\tilde \mA$, and therefore it
is possible to identify these parameters by measuring $\la
\sigma_x^1 \ra$.

Consider the case of three-qubit ($n=3$), whereby 

\begin{equation}
  \label{eq:27}
 \tilde \mA=\left[\begin{array}{cccccccccc}
-\gamma_1&      \omega_1&       0&     -\delta_1&       0&       0\\
-\omega_1&    -\gamma_1&      \delta_1&       0&       0&       0\\
0&     -\delta_1&    -\gamma_2&      \omega_2&       0&     -\delta_2\\
\delta_1&       0&     -\omega_2&    -\gamma_2&      \delta_2&       0\\
0&       0&       0&     -\delta_2&    -\gamma_3&      \omega_3\\
0&       0&      \delta_2&       0&     -\omega_3&    -\gamma_3
 \end{array}\right]. 
\end{equation}
Now, if the initial state is given by
$\ket{\psi(0)}=(\ket{0}+\ket{1})/\sqrt{2}\otimes \ket{00}$, then we have
\begin{equation}
  \label{eq:21}
  \mx_a(0)=\trans{\ma{1,&0,&0,&0,&0,&0}}.
\end{equation}
This is an example where the Laplace transform of the measurement trace $\bar{x}_1(t)$ takes the canonical form given by \erf{eq:49}, and we obtain
\begin{equation}
 \label{eq:28}
  \bar{X}_1(s)=\frac{s^5+q_4s^4+q_3s^3+q_2s^2+q_1s+q_0}
{s^6+p_5s^5+p_4s^4+p_3s^3+p_2s^2+p_1s+p_0}, \nn
\end{equation}
where the expressions for $p_k$ and $q_k$ in terms of Hamiltonian
parameters and Lindblad coefficients are :

\nobreak{
\begin{equation*}
  \begin{aligned}
p_5&=2 (\gamma_1 + \gamma_2 + \gamma_3),\\
p_4&=2 \delta_1^2 + 2 \delta_2^2 + \gamma_1^2 + 4 \gamma_1 \gamma_2 +
\gamma_2^2 + 4 (\gamma_1 + \gamma_2) \gamma_3 + \gamma_3^2 + \omega_1^2 + \omega_2^2 + \omega_3^2,\\
p_3&=2 (\gamma_1^2 \gamma_2 + \gamma_1 \gamma_2^2 + \gamma_1^2 \gamma_3 + 4 \gamma_1 \gamma_2 \gamma_3 + \gamma_2^2 \gamma_3 + \gamma_1 \gamma_3^2 + \gamma_2 \gamma_3^2+ \delta_2^2 (2 \gamma_1 + \gamma_2 +
\gamma_3) + \delta_1^2 (\gamma_1 + \gamma_2 + 2 \gamma_3) \\
&\quad+ \gamma_2 \omega_1^2 +
    \gamma_3 \omega_1^2 + \gamma_1 \omega_2^2 + \gamma_3 \omega_2^2 + (\gamma_1 + \gamma_2) \omega_3^2),\\
p_2&=\delta_1^4 + \delta_2^4 + \gamma_1^2 \gamma_2^2 + 4 \gamma_1^2
\gamma_2 \gamma_3 + 4 \gamma_1 \gamma_2^2 \gamma_3 + \gamma_1^2
\gamma_3^2 + 4 \gamma_1 \gamma_2 \gamma_3^2 + \gamma_2^2 \gamma_3^2 +
\gamma_2^2 \omega_1^2 + 4 \gamma_2 \gamma_3 \omega_1^2 + \gamma_3^2
\omega_1^2 + \gamma_1^2 \omega_2^2 + 4 \gamma_1 \gamma_3 \omega_2^2 \\
&\quad  + \gamma_3^2 \omega_2^2 + \omega_1^2 \omega_2^2 + (\gamma_1^2 + 4 \gamma_1 \gamma_2 + \gamma_2^2 + \omega_1^2 + \omega_2^2) \omega_3^2 + 2 \delta_2^2 (\gamma_1^2 + \gamma_2 \gamma_3 + 2 \gamma_1
(\gamma_2 + \gamma_3) + \omega_1^2 - \omega_2 \omega_3) \\
&\quad + 2 \delta_1^2 (\delta_2^2 + \gamma_1 \gamma_2 + 2 (\gamma_1 +
\gamma_2) \gamma_3 + \gamma_3^2 - \omega_1 \omega_2 + \omega_3^2), \\
p_1&=2 (\delta_2^4 \gamma_1 + \delta_1^4 \gamma_3 + 
   \gamma_2 \gamma_3 (\gamma_1 \gamma_2 \gamma_3 + \gamma_1^2
   (\gamma_2 + \gamma_3) + (\gamma_2 + \gamma_3) \omega_1^2) + 
   \gamma_3 (\gamma_1 (\gamma_1 + \gamma_3) + \omega_1^2) \omega_2^2  \\
&\quad + (\gamma_2 (\gamma_1 (\gamma_1 + \gamma_2) + \omega_1^2) + 
      \gamma_1 \omega_2^2) \omega_3^2 +\delta_1^2 (\delta_2^2 (\gamma_1 + \gamma_3) + 
      \gamma_3 (2 \gamma_1 \gamma_2 + (\gamma_1 + \gamma_2) \gamma_3 -
      2 \omega_1 \omega_2)\\
&\quad  + (\gamma_1 + \gamma_2) \omega_3^2) + 
   \delta_2^2 (\gamma_1^2 (\gamma_2 + \gamma_3) + (\gamma_2 +
   \gamma_3) \omega_1^2 + 2 \gamma_1 (\gamma_2 \gamma_3 - \omega_2 \omega_3))),\\
p_0&=(\delta_1^4 + 
    2 \delta_1^2 (\gamma_1 \gamma_2 - \omega_1 \omega_2) + (\gamma_1^2
    + \omega_1^2) (\gamma_2^2 + \omega_2^2))(\gamma_3^2 + 
    \omega_3^2) + 2 \delta_2^2 (\delta_1^2 (\gamma_1 \gamma_3 +
    \omega_1 \omega_3) +\delta_2^4 (\gamma_1^2 + \omega_1^2)\\
&\quad + (\gamma_1^2 + \omega_1^2) (\gamma_2 \gamma_3 - 
       \omega_2 \omega_3)),    
  \end{aligned}
\end{equation*}
}
and
\begin{equation*} 
\begin{aligned}
q_4&=\gamma_1+2\gamma_2+2\gamma_3 ,\nn \\
q_3&=\delta_1^2 + 2 \delta_2^2 + \gamma_2^2 + 4 \gamma_2 \gamma_3 +
\gamma_3^2 + 2 \gamma_1 (\gamma_2 + \gamma_3) + \omega_2^2 + \omega_3^2,\\
q_2&=2 \delta_2^2 (\gamma_1 + \gamma_2 + \gamma_3) + \delta_1^2
(\gamma_2 + 2 \gamma_3) +  2 \gamma_3 (\gamma_2 (\gamma_2 + \gamma_3) + \omega_2^2) + 2\gamma_2 \omega_3^2 +\gamma_1 (\gamma_2^2 + 4 \gamma_2 \gamma_3 + \gamma_3^2 + \omega_2^2 + \omega_3^2),\\
q_1&=\delta_2^4 + \gamma_2 \gamma_3 (\gamma_2 \gamma_3 + 2 \gamma_1
(\gamma_2 + \gamma_3)) + \gamma_3 (2 \gamma_1 + \gamma_3) \omega_2^2 + (2 \gamma_1\gamma_2 + \gamma_2^2 + \omega_2^2) \omega_3^2 + 2 \delta_2^2 (\gamma_2 \gamma_3 + \gamma_1 (\gamma_2 +
\gamma_3) - \omega_2 \omega_3)\\
&\quad  + \delta_1^2 (\delta_2^2 + 2 \gamma_2 \gamma_3 + \gamma_3^2 + \omega_3^2),\\
q_0&=\delta_2^4 \gamma_1 + 
 \delta_2^2 (\delta_1^2 \gamma_3 + 2 \gamma_1 \gamma_2 \gamma_3 - 2
 \gamma_1 \omega_2 \omega_3) + (\delta_1^2 \gamma_2 + 
    \gamma_1 (\gamma_2^2 + \omega_2^2)) (\gamma_3^2 + \omega_3^2). 
  \end{aligned}
\end{equation*}

The Hamiltonian and Lindblad parameters can be obtained by solving
these polynomial equations.

\subsection{Independent relaxation}
Next we consider another common form of Markovian decoherence, \ie 
independent relaxation \cite{Bre.Pet-2002}. This amounts to the following
Lindblad terms
\begin{equation}
  \label{eq:18}
\sum_k 2 g_k^- (\sigma_-^k\rho \sigma_+^k -\frac12
  \sigma_+^k\sigma_-^k\rho -\frac12 \rho \sigma_+^k\sigma_-^k),
\end{equation}
where $g_k^-$ is the relaxation rate of qubit $k$.

Consider for example, the two-qubit case where we directly observe $\expect{\sigma_x^1}$.  Performing the generalized filtration procedure prescribed by \erf{eq:filtration}, the accessible portion of the coherence vector is found to be $\mx_a=[\la\sigma_x^1\ra$, $\la\sigma_y^1 \ra$, $\la
\sigma_x^2\ra$, $\la\sigma_y^2 \ra$, $\la\sigma_x^1\sigma_z^1 \ra$,
$\la \sigma_y^1\sigma_z^1\ra$, $\la\sigma_z^1\sigma_x^2 \ra$, $\la
\sigma_z^1\sigma_y^2 \ra]^{\mathsf{T}}$. Comparing with \Eq{eq:26}, we see that the Lindblad relaxation term has introduced four new elements into the accessible state vector, namely, $\la \sigma_x^2\ra$, $\la\sigma_y^2 \ra$, $\la\sigma_x^1\sigma_z^1
\ra$, and $\la \sigma_y^1\sigma_z^1\ra$. The dynamical equation is
described by \erf{eq:dyn_acc}, with  
\begin{equation}
  \label{eq:136}
\tilde\mA=\left[\begin{array}{cccccccc}
 -g_1^-& \omega_1& 0& 0& 0& 0& 0& -\delta_1\\
 -\omega_1& -g_1^-& 0& 0& 0& 0& \delta_1& 0\\
 0& 0& -g_2^-& \omega_2& 0& -\delta_1& 0& 0\\
 0& 0& -\omega_2& -g_2^-& \delta_1& 0& 0& 0\\
 -2 g_2^-& 0& 0& -\delta_1& -2 g_2^--g_1^-& \omega_1& 0& 0\\
 0& -2 g_2^-& \delta_1& 0& -\omega_1& -2 g_2^--g_1^-& 0& 0\\
 0& -\delta_1& -2 g_1^-& 0& 0& 0& -2 g_1^--g_2^-& \omega_2\\
 \delta_1& 0& 0& -2 g_1^-& 0& 0& -\omega_2& -2 g_1^--g_2^-
 \end{array}\right],   
\end{equation}
and $\tilde{\mb}=0$. Note that the constant forcing vector $\tilde
\mb$ for the dynamics of the accessible set can be zero even if $\mb\neq 0$ (as in this case).
 
Suppose we prepare the initial state 
$\ket{\psi(0)}=(\ket{0}+\ket{1})/\sqrt{2}\otimes \ket{0}$, then
\begin{equation}
  \label{eq:24}
  \mx_a(0)=\ma{1,&0,&0,&0,&1,&0,&0,&0}^{\mathsf{T}}. \nn
\end{equation}
The resolvent on the left hand side of \erf{eq:realization_gen} can be computed symbolically in this case, and consequently the Laplace transform of the measurement trace $\bar{x}_1$ is
\begin{equation}
  \label{eq:37}
  \bar{X}_1(s)=\frac{s^7+q_6s^6+\cdots+q_1s+q_0}
{s^8+p_7s^7+p_6s^6+\cdots+p_1s+p_0}.
\end{equation}
The expressions for the coefficients in \Eq{eq:37} are quite involved and so we only present the most concise ones here:
\begin{equation}
  \label{eq:38}
  \begin{aligned}
q_6&=7g_1^-+8 g_2^-,\\
q_5&= 26 (g_2^-)^2+48 g_2^- g_1^-+3 \delta_1^2+2 \omega_2^2+19 (g_1^-)^2+\omega_1^2,\\
q_4&= 130 g_1^- (g_2^-)^2+17 \delta_1^2 g_2^-+5 \omega_1^2
g_1^-+10 \omega_2^2 g_1^- +12 \omega_2^2 g_2^-
 +108 (g_1^-)^2 g_2^-+44 (g_2^-)^3+18 \delta_1^2 g_1^- +25 (g_1^-)^3+4 \omega_1^2 g_2^-, \nn
  \end{aligned}
\end{equation}
and
\begin{equation}
  \label{eq:39}
  \begin{aligned}
p_7&=8g_1^-+8g_2^-, \\
p_6&=2 \omega_2^2+26 (g_1^-)^2+26 (g_2^-)^2
+56 g_2^- g_1^-+2 \omega_1^2+4 \delta_1^2.
  \end{aligned}
\end{equation}
These five equations and the realization formed from the measurement data can be used to solve for the five unknown parameters in this model.

To treat a case where $\tilde \mb \neq 0$, we now alter the setup to
consider direct measurement of a time trace of $\bar z_1 \equiv
\expect{\sigma_z^1}$. In this case, with the Hamiltonian given in \erf{eq:ham_eg} with $n=2$ and independent relaxation as prescribed in \erf{eq:18}, the accessible set becomes $\bar{G} = \{\sigma_z^1, \sigma_z^2,
\sigma_x^1\sigma_x^2, \sigma_x^1\sigma_y^1, \sigma_y^1\sigma_x^2,
\sigma_y^1\sigma_y^2\}$. The
dynamics of $\mx_a$ is determined by \erf{eq:dyn_acc} with
\begin{equation}
  \begin{aligned}
&\tilde \mA=\ma{-2 g_1^-& 0& 0& \delta_1& -\delta_1& 0\\
  0& -2 g_2^-& 0& -\delta_1& \delta_1& 0\\
  0& 0& -g_s^-& \omega_2& \omega_1& 0\\
  -\delta_1& \delta_1& -\omega_2& -g_s^-& 0& \omega_1\\
  \delta_1& -\delta_1& -\omega_1& 0& -g_s^-& \omega_2\\
  0& 0& 0& -\omega_1& -\omega_2& -g_s^-},\\
&\tilde \mb=\ma{-g_1^-,&-g_2^-,&0,&0,&0,&0}^\mathsf{T}, \nn
  \end{aligned}
\end{equation}
where $g_s^-=g_1^-+g_2^-$.
In this basis $\mC=\left[ 1, 0, 0, 0, 0, 0 \right]$. Let the initial
state be $\ket{\psi(0)}=(\ket{0}+\ket{1})/\sqrt{2}\otimes \ket{0}$ as before, in which case
\begin{equation}
  \label{eq:45}
   \mx_a(0)=\ma{0,&1,&0,&0,&0,&0}^\mathsf{T}. \nn   
\end{equation}
Again, we can symbolically calculate the resolvents in this case, and obtain the Laplace transform of the measurement trace $\bar{z}_1$ as
\begin{equation}
  \label{eq:16}
  \begin{aligned}
\bar{Z}_1(s)&=\mC(s\mI-\mA)^{-1}\mx_a(0)+\mC(s\mI-\mA)^{-1}\mb/s\\
&=\frac{q_3 s^3+q_2s^2+q_1s+q_0}
{s^5+p_4s^4+p_3s^3+p_2s^2+p_1s},
  \end{aligned}
\end{equation}
with
\begin{equation}
  \label{eq:50}
    \begin{aligned}
q_3&=-g_1^-,\\
q_2&=2\delta_1^2-2(g_1^-)^2-4g_1^-g_2^-,\\
q_1&=-g_1^-\left((g_1^-)^2+6g_1^-g_2^-+5(g_2^-)^2
 +\omega_d^2\right ),\\
q_0&=-2\delta_1^2(g_1^-+g_2^-)^2-2g_1^-g_2^-( (g_1^-+g_2^-)^2
+\omega_d^2),
  \end{aligned}
\end{equation}
and
\begin{equation}
\label{eq:51}
\begin{aligned}
p_4&=4(g_1^-+g_2^-),\\
p_3&=4\delta_1^2+5(g_1^-)^2+14g_1^-g_2^-
 +5(g_2^-)^2+\omega_d^2,\\
p_2&=2(g_1^-+g_2^-)(4\delta_1^2+(g_1^-)^2+6g_1^-g_2^-+(g_2^-)^2
+\omega_d^2), \\
p_1&=4\delta_1^2 (g_1^-+g_2^-)^2+4g_1^-g_2^-
( (g_1^-+g_2^-)^2+\omega_d^2),
\end{aligned}
\end{equation}
where $\omega_d=\omega_1-\omega_2$.  An interesting aspect of this
example is that from \Eqs{eq:50} and~\eqref{eq:51} we can
identify $g_1^-$, $g_2^-$, and $\delta_1$, but only
$\omega_1-\omega_2$. The individual transition energies of the qubits
do not influence the measurement trace, and only the their difference
does.

\end{widetext}

\bibliographystyle{apsrev}
\bibliography{refs}

\end{document}